\documentclass[traditabstract]{aa} 

\usepackage{graphicx}
\usepackage{natbib}
\usepackage{txfonts}
\begin{document}
   \title{Kn\,26, a New Quadrupolar Planetary Nebula\thanks{
Based on observations made with 
the Nordic Optical Telescope (NOT) and the William Herschel Telescope (WHT) 
on the island of La Palma in the Spanish Observatorio del Roque de los 
Muchachos of the Instituto de Astrof\'\i sica de Canarias (IAC), 
the 2.1-m telescope of the Observatorio Astron\'omico Nacional at the 
Sierra de San Pedro M\'artir (OAN-SPM), and 
the 1.5-m telescope at the Observatorio de Sierra Nevada 
(OSN), Granada, Spain.  
NOT is operated jointly by Denmark, Finland, Iceland, Norway, and Sweden. 
WHT is operated by the Isaac Newton Group.  
The 2.1-m telescope at the OAN-SPM is a national facility operated 
by the Instituto de Astronom\'\i a of the Universidad Nacional 
Aut\'onoma de M\'exico.  
The 1.5-m telescope at the OSN is operated by the Instituto de 
Astrof\'\i sica de Andaluc\'\i a (IAA).
The data presented here were obtained in part with ALFOSC, which is 
provided by the IAA under a joint agreement with the University of 
Copenhagen and NOTSA.
}
} 

   \author{M.A. Guerrero
          \inst{1}
          \and
          L.F.\ Miranda
          \inst{2,3}
          \and
          G.\ Ramos-Larios
          \inst{4}
          \and
          R.\ V\'azquez
          \inst{5}
          }

\institute{
Instituto de Astrof\'{\i}sica de Andaluc\'{\i}a (IAA-CSIC), 
Glorieta de la Astronom\'{\i}a s/n, E-18008 Granada, Spain 
\and
Departamento de F\'\i sica Aplicada, Facultade de Ciencias, 
Campus Lagoas-Marcosende s/n, Universidade de Vigo, E-36310, Vigo, Spain 
\and
Consejo Superior de Investigaciones Cient\'\i ficas (CSIC), 
c/ Serrano 117, E-28006 Madrid, Spain
\and
Instituto de Astronom\'{\i}a y Meteorolog\'{\i}a, 
Av.\ Vallarta No.\ 2602, Col.\ Arcos Vallarta, C.P. 44130 Guadalajara, 
Jalisco, Mexico
\and
Instituto de Astronom\'{\i}a, Universidad Nacional Aut\'onoma de M\'exico, 
Apdo. Postal 877, 22800 Ensenada, B.C., Mexico \\ 
\email{mar@iaa.es, lfm@iaa.es, gerardo@astro.iam.udg.mx, vazquez@astro.unam.mx}
}

   \date{Received sometime; accepted later}

\abstract{
Once classified as an emission line source, the planetary nebula (PN) nature 
of the source Kn\,26 has been only recently recognized in digital sky surveys.  
To investigate the spectral properties and spatio-kinematical structure 
of Kn\,26, we have obtained high spatial-resolution optical and near-IR 
narrow-band images, high-dispersion long-slit echelle spectra, and 
intermediate-resolution spectroscopic observations. 
The new data reveal an hourglass morphology typical of bipolar PNe.  
A detailed analysis of its morphology and kinematics discloses the presence 
of a second pair of bipolar lobes, making Kn\,26 a new member of the subclass 
of quadrupolar PNe. 
The time-lap between the ejection of the two pairs of bipolar lobes 
is much smaller than their dynamical ages, implying a rapid change 
of the preferential direction of the central engine.  
The chemical composition of Kn\,26 is particularly unusual among PNe, 
with a low N/O ratio (as of type~II PNe) and a high helium abundance 
(as of type~I PNe), although not atypical among symbiotic stars.  
Such an anomalous chemical composition may have resulted from 
the curtail of the time in the Asymptotic Giant Branch by the 
evolution of the progenitor star through a common envelope 
phase.  
}


   \maketitle
%

\section{Introduction}

In recent years, narrow-band optical surveys of the Galaxy and near- and 
mid-IR mapping of the sky have incessantly increased the population of 
known Galactic planetary nebulae (PNe) and their immediate precursors, 
post-AGB stars and proto-PNe 
\citep[e.g.,][]{Parker_etal06,Setal06,Miszalski_etal08,Viironen_etal09}.  
This observational effort has allowed us a better assessment of the 
role of PNe in the Galaxy chemical enrichment and on the processes 
of PNe formation and evolution.  
Incidentally, these surveys have revealed a number of PNe with very 
peculiar morphologies, physical structures and evolutionary situations 
\citep[e.g.,][]{Mampaso_etal06,Miszalski_etal11}.
These new objects are providing interesting case studies to 
investigate the complexity of the PN phenomenon.

Using existing digital sky surveys such as the POSS-I and POSS-II surveys, 
\citet{Jacoby_etal10} presented Kn\,26, a bipolar PN candidate previously 
known as the emission line source Lan\,384 \citep{Lanning00,Eracleous_etal02}.  
An inspection of the narrow-band H$\alpha$ image of Kn\,26 
presented by \citet{Jacoby_etal10} suggests a bipolar morphology 
with an intriguing S-shaped point-symmetric structure, whereas the 
optical spectroscopy presented by \citet{Eracleous_etal02} supports 
its classification as a PN.

To confirm the PN nature of Kn\,26 and to investigate its morphology, 
kinematics, physical structure and physical conditions and chemical 
abundances, we have obtained high spatial-resolution optical and 
near-IR narrow-band images of this nebula in conjunction with 
intermediate-dispersion and echelle long-slit spectroscopic observations.  
The analyses of these data presented in this paper allow us to 
conclude that Kn\,26 is a true PN (PN\,G084.7--08.0, following the 
standard rules of nomenclature for PNe), whose spatio-kinematical 
properties make a new member of the quadrupolar class of PNe 
\citep{Manchado_etal96}.  
We next describe the observations in Sect.\ 2 
and provide the main results in Sect.\ 3.  
These are discussed in Sect.\ 4 and summarized in Sect.\ 5.

\begin{table*}
\caption{Properties of the Narrow-band Filters}
\label{tab.filt}
\centering
\begin{tabular}{lrrc|lrrc}
\hline\hline
\multicolumn{1}{l}{Optical Filter}    & 
\multicolumn{1}{c}{$\lambda_{\rm c}$}  & 
\multicolumn{1}{c}{$\Delta\lambda$}   & 
\multicolumn{1}{l}{Transmission peak} & 
\multicolumn{1}{l}{Near-IR Filter}    & 
\multicolumn{1}{c}{$\lambda_{\rm c}$}  & 
\multicolumn{1}{c}{$\Delta\lambda$}   & 
\multicolumn{1}{l}{Transmission peak} \\
\multicolumn{1}{c}{}                  & 
\multicolumn{1}{c}{(\AA)}             & 
\multicolumn{1}{c}{(\AA)}             &        
\multicolumn{1}{c}{(\%)}              & 
\multicolumn{1}{c}{}                  & 
\multicolumn{1}{c}{($\mu$m)}          & 
\multicolumn{1}{c}{($\mu$m)}          & 
\multicolumn{1}{c}{(\%)}              \\ 
\hline
{[}O~{\sc iii}] & 5007 & 30~ & 77 & H$_2$         & 2.122 & 0.032 & 70~ \\ 
H$\alpha$       & 6567 &  8~ & 60 & Br$\gamma$    & 2.166 & 0.032 & 73~ \\
{[}N~{\sc ii}]  & 6588 &  9~ & 62 & $K$ continuum & 2.270 & 0.034 & 72~ \\
\hline
\end{tabular}
\end{table*}

\section{Observations}

\subsection{Narrow-band imaging}

\begin{figure*}[t]
\begin{center}
 \includegraphics[width=11.0cm,bb=0 -8 656 374]{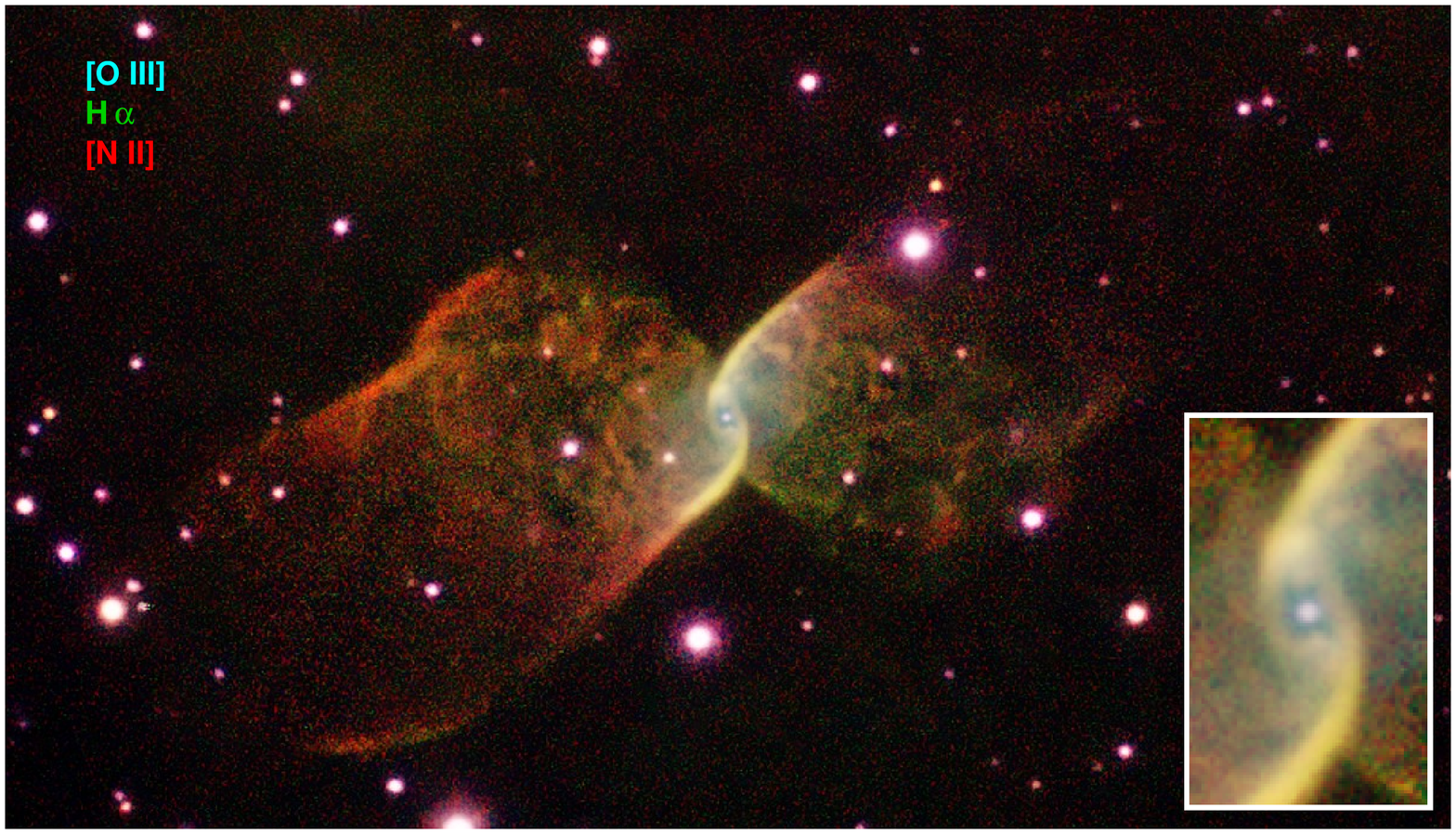}
 \includegraphics[width=11.0cm,bb=0 -8 656 374]{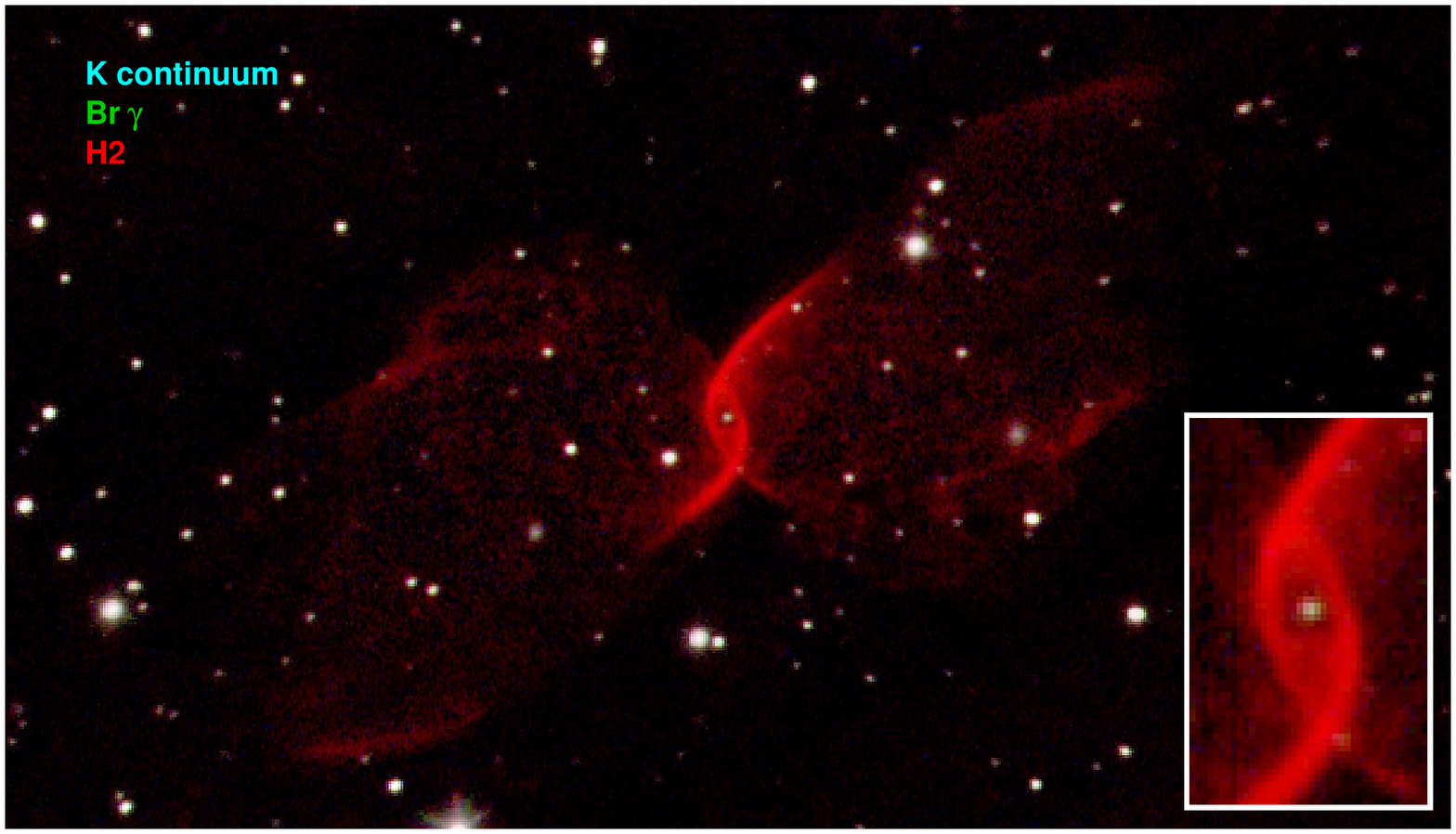}
 \includegraphics[width=11.0cm]{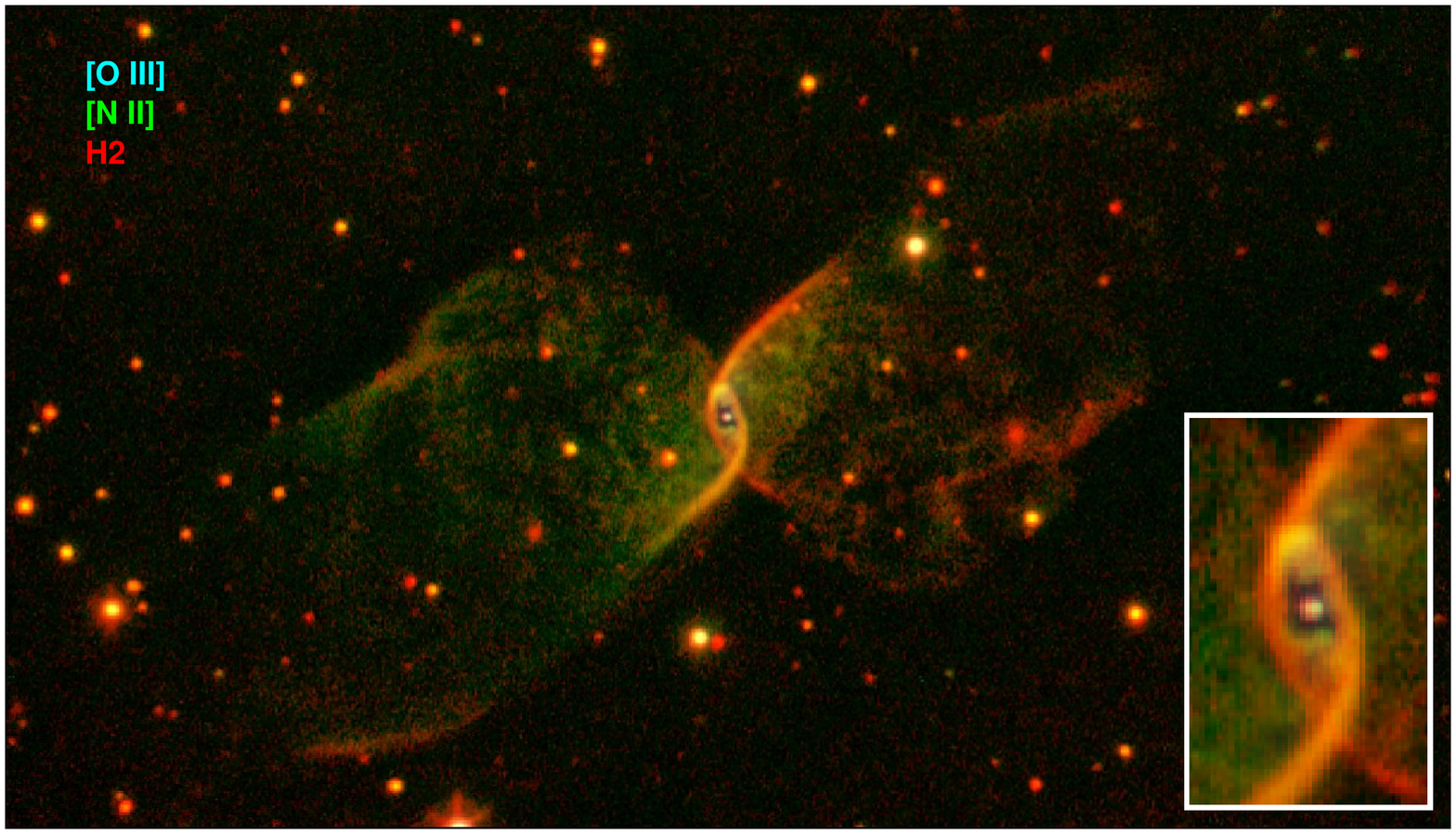}
\caption{
Color-composite optical ({\it top}), near-IR ({\it center}), and 
optical/near-IR ({\it bottom}) narrow-band pictures of Kn\,26.  
The narrow-band filters and colors assigned to each picture are 
labeled on them.  
The FoV is 150\arcsec$\times$85\arcsec, whereas the insets show in 
greater detail the innermost 8\farcs5$\times$14\farcs5 nebular regions.  
In all pictures north is up, east to the left.  
}
   \label{img1}  
\end{center}
\end{figure*}

Narrow-band H$\alpha$, [N~{\sc ii}] $\lambda$6583, and [O~{\sc iii}] 
$\lambda$5007 images of Kn\,26 have been acquired on 2009 June 21 using 
ALFOSC (Andalucia Faint Object Spectrograph and Camera) at the 2.56m 
Nordic Optical Telescope (NOT) of the Observatorio de Roque de los 
Muchachos (ORM, La Palma, Spain).  
The central wavelength ($\lambda_{\rm c}$), bandwidth ($\Delta\lambda$), 
and transmission peaks of these filters are provided in Table~\ref{tab.filt}. 
The EEV 2048$\times$2048 CCD with pixel size 13.5 $\mu$m was used as 
detector and the exposure time was 900 s for each filter.  
The images have a plate scale of 0$\farcs$184 pixel$^{-1}$, a field of view 
(FoV) 6$\farcm$3$\times$6$\farcm$3, and a spatial resolution of 0$\farcs$7, 
as determined from the FWHM of stars in the FoV.  
The data were bias-subtracted and flat-fielded by twilight flats using 
standard IRAF\footnote{
IRAF is distributed by the National Optical Astronomy Observatory, 
which is operated by the Association of Universities for Research 
in Astronomy, Inc., under cooperative agreement with the National 
Science Foundation.  
}
V2.14.1 routines.
Figure~\ref{img1}-{\it top} shows a color-composite picture of the 
optical narrow-band images of Kn\,26.

Narrow-band H$_2$ 2.1218 $\mu$m, Br$\gamma$ 2.1658 $\mu$m, and 
$K$ continuum at 2.270 $\mu$m images of Kn\,26 were obtained on 
2010 June 27 using LIRIS \citep[Long-Slit Intermediate Resolution 
Infrared Spectrograph,][]{Pulido_etal03} at the Cassegrain focus 
of the 4.2m William Herschel Telescope (\emph{WHT}) at the ORM.  
As for the optical filters, the central wavelength, bandwidth, and 
transmission peak of these filters are listed in Table~\ref{tab.filt}.  
The detector was a 1k$\times$1k HAWAII array with plate scale 
0$\farcs$25 pixel$^{-1}$ and the FoV 4$\farcm$3$\times$4$\farcm$3. 
We obtained series of 4 exposures with integration time 60 s on each 
filter, for total effective exposure times of 720 s for H$_2$ and 
Br$\gamma$, and 480 s for $K$ continuum.  
For each series of 4 exposures, the nebula was placed at the center 
of each quadrant of the detector to acquire simultaneously the object 
and, by combining directly the 4 exposures, the sky.  
The data reduction was carried out using the dedicated software LIRISDR 
(LIRIS Data Reduction package), a pipeline for the automatic reduction 
of near-IR data developed within the IRAF environment.  
The reduction performed by LIRISDR includes standard and additional 
non-standard steps such as bad pixel mapping, cross-talk correction, 
flat-fielding, sky subtraction, removal of reset anomaly effect, 
field distortion correction, and final image shift and co-addition. 
Figure~\ref{img1}-{\it center} shows a color-composite picture of the 
near-IR narrow-band images of Kn\,26.  
The lack of nebular continuum emission and the brighter emission in H$_2$ 
with respect to Br$\gamma$ results in the red appearance of the nebula 
in this picture.  
The spatial resolution, as determined from the FWHM of stars 
in the FoV, is $\approx$0$\farcs$8.

In addition, we have registered the optical and near-IR images to compare 
the emission in the H$_2$, [N~{\sc ii}], and [O~{\sc iii}] emission lines.  
The color composite picture is shown in Figure~\ref{img1}-{\it bottom}.

\subsection{Spectroscopic observations}

Intermediate-resolution long-slit spectra of Kn\,26 were obtained on 2011 
October 5, using the ALBIREO spectrograph at the 1.5 m telescope of the 
Observatorio de Sierra Nevada (OSN), Granada, Spain.  
A Marconi 2048$\times$2048 CCD was used as a detector, in conjunction 
with a 400 l~mm$^{-1}$ grating blazed at 5500 \AA. 
The slit length was $\approx$6\arcmin\ and its width was set at 50 $\mu$m  
($\equiv$2.5\arcsec) to match the seeing during the observations.  
The binning 2$\times$2 of the detector implied plate and spectral 
scales of 0\farcs30~pix$^{-1}$ and 1.89~\AA~pix$^{-1}$, respectively. 
The spectral resolution was $\approx$4.7 \AA, the wavelength 
uncertainty $\approx$1 \AA, and the spectral range covered 
3600--7200 \AA.

Two positions with exposures of 1800 seconds were obtained with the slit 
centered on the central star and oriented along the position angles (P.A.) 
112\degr\ and 147\degr, i.e., along the axis of the major bipolar lobes 
and along the bright S-shaped region.  
The observations were flux calibrated using spectra of the spectrophotometric 
standard stars G~191-B2B and Hiltner~600 acquired on the same night.  
All spectra were bias-subtracted, flat-fielded, wavelength, and flux 
calibrated following standard IRAF procedures.

Long-slit high dispersion spectroscopy on the H$\alpha$ and [N~{\sc ii}] 
$\lambda$6583 lines of Kn\,26 has been acquired on 2010 June 13 using the 
Manchester Echelle Spectrometer \citep[MES,][]{Meaburn_etal03} mounted on 
the 2.1\,m (f/7.5) telescope at the Observatorio Astron\'omico Nacional 
de San Pedro M\'artir (OAN-SPM, Mexico).  
The $2048\times2048$ Thomson CCD with a pixel size of $15\mu$m was used, 
resulting a plate scale of $0\farcs352\,{\rm pixel}^{-1}$ and a dispersion 
of 0.06\,{\AA}\,pixel$^{-1}$. 
The 2$^{\prime\prime}$ wide slit was set across the central star and 
oriented along the axes of the major bipolar lobes (P.A.=110\degr) 
and minor bipolar lobes (P.A.=65\degr) with on-chip binning 1$\times$1 
and 2$\times$2 and spectral resolutions $\approx$6 km~s$^{-1}$ and 
$\approx$12 km~s$^{-1}$, respectively. 
The spectra were wavelength calibrated with a Th-Ar arc lamp to an 
accuracy of $\pm1$\,km\,s$^{-1}$.

\section{Results}

\subsection{Morphology}

\begin{figure}[t]
\begin{center}
\includegraphics[width=1.0\columnwidth]{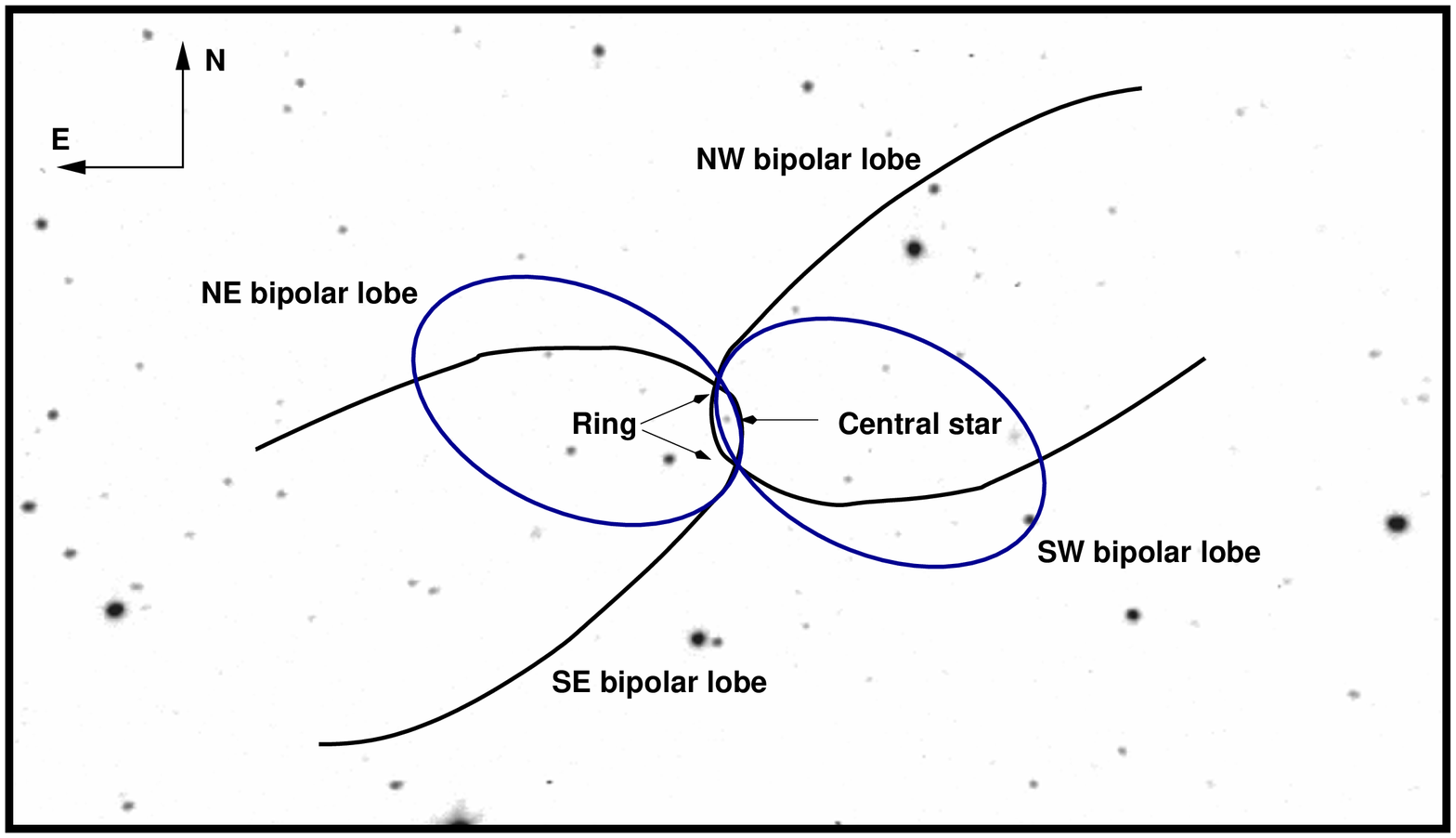}
\caption{
Schematic drawing of the two pairs of bipolar lobes of Kn\,26 with the 
different morphological components labeled on it.  
The eastern minor and major bipolar lobes recede from us, whereas the 
western lobes approach to us.  
}
   \label{sketch}  
\end{center}
\end{figure}

The images of Kn\,26 in Figure~\ref{img1} reveal the following 
morphological components:  
(1) the major bipolar lobes, a pair of large bipolar lobes extending 
$\approx$110\arcsec\ along PA $\approx$110\degr; 
(2) the minor bipolar lobes, a pair of small bipolar lobes extending 
$\approx$75\arcsec\ along PA $\approx$75\degr; and 
(3) a central elliptical ring.  
These components, marked on the sketch of Kn\,26 in Figure~\ref{sketch}, 
are described in more detail next.

The major bipolar lobes, very prominent in [N~{\sc ii}] and H$_2$, have 
open ends, becoming very faint at large distances from the nebular center. 
Their inner regions show a clear point-symmetric brightness distribution 
defined by two arcs that trace the central ring and the edges of the 
lobes in these innermost regions.  
This same point-symmetric intensity distribution is present at other locations 
of the lobes, very particularly in the H$_2$ image, such as the bars located 
36\arcsec--54\arcsec\ from the nebula center that trace the southern edge of 
the SE bipolar lobe and northern edge of the NW lobe, and the regions at 
36\arcsec\ and PA $\approx$75\degr\ and 255\degr\ coincident with the polar 
caps of the minor bipolar lobes that define the northern and southern edges of 
the SE and NW bipolar lobes, respectively. 
The H$\alpha$ image presents similar structures as those observed in 
[N~{\sc ii}], whereas in the [O~{\sc iii}] image the point-symmetric arcs 
are observed as a high-excitation region (blue in Fig.~\ref{img1}-{\it top}) 
along the major nebular axis with an extent of $\approx$5\arcsec\ at both 
sides of the star at the center of the nebula.

The minor bipolar lobes have elliptical shape (Fig.~\ref{sketch}) and 
are closed, at variance with the major bipolar lobes.  
The NE lobe has a maximum extent from the center of 31\farcs6, 
while the SW lobe reaches up to 34\farcs7.  
The polar regions of these lobes are particularly bright, especially for 
the NE lobe.  
As for the major bipolar lobes, the inner regions of the minor bipolar 
lobes share the arcs that define the central ring.

The central ring has an elliptical shape, with its minor axis along PA 
$\approx$100\degr, i.e., similar but no completely coincident with the 
orientation of the major bipolar lobes.  
The size of the ring is 
8\farcs3$\times$2\farcs9 in H$_2$, 
7\farcs7$\times$2\farcs3 in [N~{\sc ii}], and 
7\farcs4$\times$2\farcs1 in H$\alpha$.  
This ring is formed by two arcs that cross at the tips of the major axis 
and extend along the edges of the bipolar lobes.  
The ring formed by these two arcs is not empty, but complex structures 
are detected inside this ring in different images, particularly two 
[N~{\sc ii}] and [O~{\sc iii}] bright knots observed at both sides of 
the star at the center.

Figure~\ref{img1}-{\it top} provides information on the spatial variations 
of the excitation in Kn\,26.  
The major bipolar lobes present low-excitation and are dominated by 
[N~{\sc ii}] emission.  
The H$_2$ emission is particularly bright in the point-symmetric regions.  
In the minor bipolar lobes, the H$\alpha$ to [N~{\sc ii}] line ratio 
is larger than in the major bipolar lobes (the green color in 
Fig.~\ref{img1}-{\it top}). 
Finally, higher excitation material, as revealed by the [O~{\sc iii}] emission, 
is concentrated at the center of the nebula, in a region $\approx$9\arcsec\ in 
size at the center of the nebula that extends along the axis of the major 
bipolar lobes.

Figure~\ref{img1}-{\it bottom} provides information about the relative 
distribution of molecular (H$_2$) and ionized material ([N~{\sc ii}] and 
[O~{\sc iii}]) in Kn\,26. 
The H$_2$ emission delineates the [N~{\sc ii}] emission, which is always 
inside the nebula.  
H$_2$ is particularly bright in the point-symmetric regions of the nebula, 
namely, in the bright point-symmetric arcs and central ring, in the two 
linear features at the south and north ends of the eastern and western major 
bipolar lobes, respectively, and at the polar regions of the minor bipolar 
lobes.

\subsection{Kinematics}

\begin{figure*}[t]
\begin{center}
\includegraphics[width=16cm]{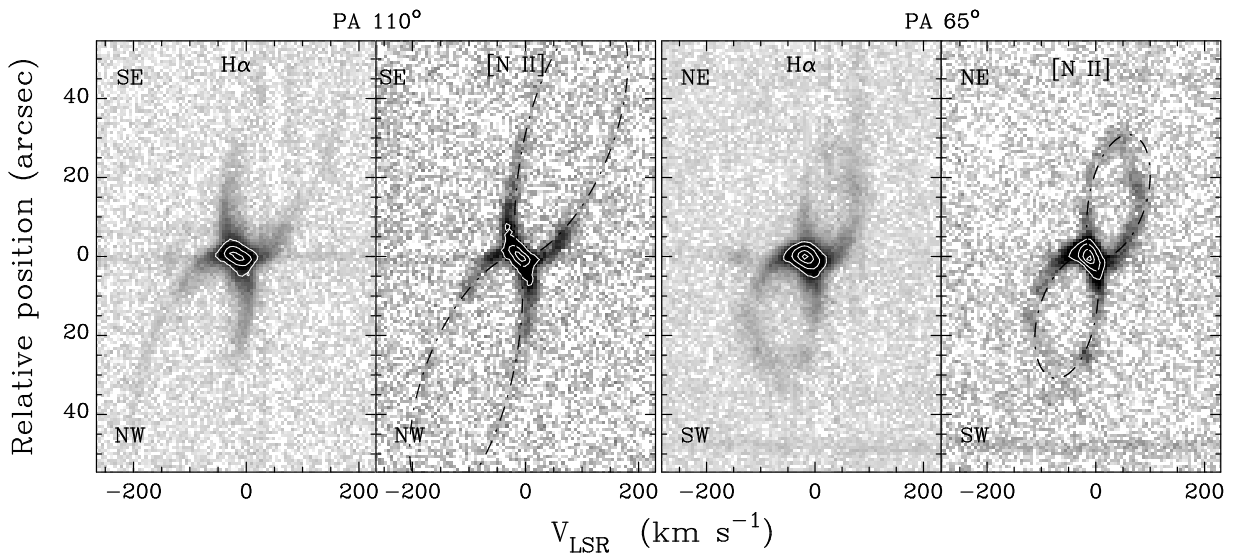}
\caption{
Position-velocity (PV) maps in the H$\alpha$ and [N~{\sc ii}] $\lambda$6583 
emission lines along the two pairs of bipolar lobes at P.A.'s 110\degr\ 
(major bipolar lobes) and 65\degr\ (minor bipolar lobes). 
The levels of the contours overlaid on the PV maps have been selected 
to emphasize the kinematical structure of the emission in the brightest 
central regions.  
The dash-dotted lines overlaid on the [N~{\sc ii}] PV maps correspond 
to the synthetic emission lines derived from our simultaneous fit to 
the morphology and kinematics of the two pairs of bipolar lobes.  
}
   \label{PV}  
\end{center}
\end{figure*}

The position-velocity maps (PV maps) of the H$\alpha$ and [N~{\sc ii}] 
$\lambda$6583 emission lines presented in Figure~\ref{PV} clearly reveal 
bipolar kinematics along both the major (PA~100\degr) and minor 
(PA~65\degr) lobes. 
The two pairs of bipolar lobes have different kinematical properties, 
but in both cases the eastern lobe recedes from us, whereas the western 
lobe moves away from the systemic velocity $v_{\rm LSR}\approx-10$ km~s$^{-1}$ 
derived from our high-dispersion spectroscopic observations.  
The major bipolar lobes, registered by the slit along PA=110\degr, are 
confirmed to be open, with velocity split between the approaching and 
receding components increasing with distance to the central star.  
On the other hand, the minor bipolar lobes, registered by the slit along 
PA=65\degr, are closed and the velocity split also shows a smooth increase 
that suddenly breaks at a distance $\approx$13\arcsec\ from the central star, 
where the approaching and receding sides of the lobes converge rather 
abruptly.

The distortion of the velocity field of the minor bipolar lobes with respect 
to a classical hour-glass expansion hints at their interaction with the major 
bipolar lobes.  
The brightening of the polar caps of the minor bipolar lobes 
and the diffuse appearance of the H$\alpha$ line in the PV map 
in these regions further support this interaction.  
We note that the minor and major bipolar lobes overlap on the regions covered 
by the slit along PA=65\degr, however, only emission from one system of bipolar 
lobes is detected in this PV map.  
Apparently, the two pairs of bipolar lobes become a unique 
structure wherever they overlap, i.e., they do not intersect.

Finally, we would like to note that the tilt of the brightest regions of the 
H$\alpha$ and [N~{\sc ii}] emission lines in the PV maps seems different, 
with the contours of the [N~{\sc ii}] lines in Fig.~\ref{PV} having larger 
inclinations than those of the H$\alpha$ lines.  
It is unclear whether this is an effect of the larger thermal 
broadening of the H$\alpha$ line, an additional contribution 
from a broad H$\alpha$ line at the location of the central 
star, or the detection of emission from the bright [N~{\sc ii}] 
knots by the side of the central star.

\subsection{Physical Model}
\label{model.sect}

We have used the visualization and modeling tool SHAPE \citep{Steffen11} to 
fit simultaneously the morphology shown in the [N~{\sc ii}] image and the 
kinematics displayed in the PV maps of the two pairs of expanding bipolar 
lobes of Kn\,26 by adopting the simple model introduced by \citet{SU85} to 
describe the structure and expansion of the nebula around the symbiotic 
Mira R~Aqr, 
\begin{equation}
v_{\rm exp}(\varphi) = 
v_{\rm e} + (v_{\rm p} - v_{\rm e}) \times {\mid{\sin \varphi}\mid}^{\alpha},
\end{equation}
where $\varphi$ is the latitude angle, varying from 0\degr\ at the 
equator to 90\degr\ at the poles, $v_{\rm e}$ and $v_{\rm p}$ are the 
polar and equatorial velocities, respectively, and $\alpha$ is a 
parameter that determines the shape of the bipolar lobes.

We have applied this model to the inner bipolar lobes and derived an 
inclination angle of 55\degr\ with respect to the line of sight, and 
polar and equatorial velocities of 160$\pm$15 km~s$^{-1}$ and $\sim$10 
km~s$^{-1}$, respectively.  
The quality of the fit is shown by the line over-plotted on the 
[N~{\sc ii}] echellogram at PA~65\degr\ (Fig.~\ref{PV}).  
As for the major bipolar lobes, a similar fit is difficult because the 
bipolar lobes are opened, thus providing little constraint on the polar 
velocity.  
A close inspection of the faintest emission from these bipolar lobes 
in the direct images and echellogram at PA~110\degr\ suggests that 
the lobes may close at a distance $\sim$63\arcsec\ from the central 
star.  
Assuming this size for the major bipolar lobes, the best-fit is 
achieved for an inclination angle also of 55\degr, and polar and 
equatorial velocities of 300$\pm$20 km~s$^{-1}$ and $\sim$12 km~s$^{-1}$, 
respectively.  
The best-fit model provides a reasonable fit of the [N~{\sc ii}] 
echellogram at PA~110\degr\ (Fig.~\ref{PV}) and lobe width, 
whereas the lobe length is uncertain.

For the minor lobes, the kinematical age of its model at a distance of 
$d$ kpc is (1125$\pm$100)$\times d$ yr, whereas for the major lobes 
only a lower limit $\gtrsim$(1150$\pm$100)$\times d$ yr can be derived.  
We note that, for the radial velocity $v_{\rm LSR}{\approx}-10$ km~s$^{-1}$, 
the Galactic coordinates of Kn\,26 ($l$=84\fdg67, $b$=--7\fdg96) 
imply a distance of 1 kpc for pure circular rotation and for a 
flat rotation curve.  
It is thus very likely that the kinematical ages of both pairs of 
bipolar lobes are in the range 1000--1300 yr.

\subsection{Physical Conditions and Chemical Abundances}

\begin{figure*}[t]
\begin{center}
\includegraphics[bb=38 386 585 710,width=0.90\linewidth]{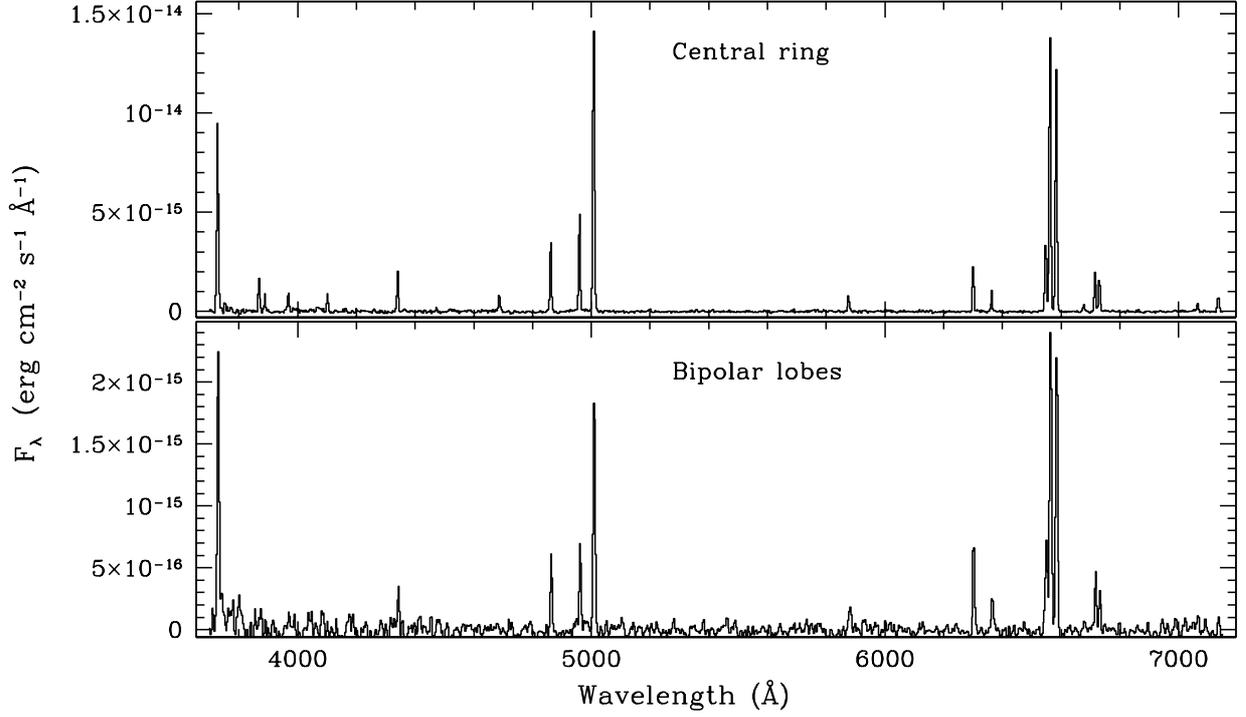}
\caption{
One-dimensional spectra of the central ring ({\it top}) and bipolar 
lobes ({\it bottom}) of Kn\,26.  
}
   \label{fig.alb}  
\end{center}
\end{figure*}

One-dimensional spectra of the central ring and bipolar lobes of Kn\,26 
have been extracted from the long-slit intermediate-dispersion ALBIREO 
spectra (Figure~\ref{fig.alb}).  
These spectra include multiple oxygen, neon, sulfur, nitrogen, and 
argon forbidden lines, as well as hydrogen and helium recombination 
lines.  
The intrinsic line intensity ratios scaled to an arbitrary H$\beta$ 
flux of 100 are listed in Table~\ref{tab.flux} where the \citet{CCM89} 
extinction law has been used to deredden the measured line intensity 
ratios using the logarithmic extinction coefficient 
$c({\rm H}\beta)$=0.30$\pm$0.04 derived from the observed H$\alpha$/H$\beta$ 
ratio for case B recombination.  
This value of the logarithmic extinction coefficient is coincident with 
the reddening of $E(B-V) =$ 0.2 derived by \citet{Eracleous_etal02}.

The line ratios listed in Table~\ref{tab.flux} for the central ring are 
generally consistent with those presented by \citet{Eracleous_etal02}, but we 
note that the intensity ratio of the [S~{\sc ii}] $\lambda\lambda$6716,6731 
lines in our spectrum is $\approx$6 times lower.  
An inspection of the spectrum of Lan\,384 presented by 
\citet{Eracleous_etal02} suggests that the emission 
line strengths for the [S~{\sc ii}] $\lambda\lambda$6716,6731 
lines listed in their Table~3 are erroneous.  
We also remark that the [O~{\sc iii}] $\lambda$5007/H$\beta$ and 
He~{\sc ii} $\lambda$4686/H$\beta$ intrinsic intensity ratios 
that can be derived from the emission line strengths listed in 
Table~3 of \citet{Eracleous_etal02}, $\approx$5.0 and $\approx$0.6, 
respectively, imply higher excitation than that of our spectrum 
of the central ring of Kn\,26 ($\approx$4.0 and $\approx$0.26, 
respectively).  
These differences reflect the higher excitation of the central 
regions of Kn\,26 along the axis of the major bipolar lobes 
(see Fig.~\ref{img1}-{\it top}) which were preferentially 
registered by the long-slit used by \citet{Eracleous_etal02} 
in their spectroscopic observations.  
At any rate, the relatively high [O~{\sc iii}] $\lambda$5007/H$\beta$ 
and He~{\sc ii} $\lambda$4686/H$\beta$ intrinsic intensity ratios found 
in both studies are typical of PNe rather than H~{\sc ii} regions.

The nebular analysis software ANNEB \citep{Olguin_etal11} which integrates 
the NEBULAR package of IRAF/STSDAS \citep{SD95} was further used to derive 
the physical conditions and nebular abundances of Kn\,26 listed in 
Table~\ref{tab.chem}.  
The NEBULAR package uses a 5-level atom approximation to compute 
the electron temperature, density, and ionic abundances of nebular 
low density gas for the most important heavy atoms.  
As for the abundances of ions of helium, these were derived following the 
method described by \citet{VKL98}, including a correction of collisional 
effects \citep{Clegg87,BSS99}. 
Since only one or a few number of ionization stages of heavy elements 
are observed in the optical spectrum, ionization correction factors 
have been adopted to compute the elemental abundances \citep{KB94}.

The electron density-sensitive ratio [S~{\sc ii}] $\lambda$6716/[S~{\sc ii}] 
$\lambda$6731 implies a low density for the nebula, $\approx$360 cm$^{-3}$.  
Such a low electron density, typical of the bipolar lobes of PNe, supports 
the idea that the apparent ring around the central star is not a real, dense 
ring, but an effect caused by the projection of the bipolar lobe edges.  
The electron temperature derived from the [N~{\sc ii}] emission lines, 
$\approx$9900~K, is notably lower than the temperature of 15000~K 
derived by \citet{Eracleous_etal02} from the [O~{\sc iii}] emission 
lines\footnote{
Unfortunately we cannot reproduce the determination of this temperature 
because the notable brightness of the Hg~{\sc i} $\lambda$4458~\AA\ sky 
line at OSN, combined with the $\approx$4.7~\AA\ spectral resolution of 
our spectra, precludes us from an accurate measurement of the intensity 
of the coronal line of [O~{\sc iii}] at 4363~\AA.
}.  
Since our slit maps regions of lower excitation than that used by 
\citet{Eracleous_etal02}, the temperature of 9900~K has been used 
for the determination of ionic abundances of the central region 
of Kn\,26 listed in Table~\ref{tab.chem}.  
Compared to other PNe, the chemical abundances of Kn\,26 place it among 
the type I PNe for its high He/H ratio, but its N/O ratio is low for PNe 
of this type and will designate it as a type II PNe\footnote{
A type III classification is precluded because the peculiar velocity 
of Kn\,26, i.e., the difference between its radial velocity and that 
expected on the basis of a pure circular motion around the Galactic 
center for sensible distances in the range 1--6 kpc, is smaller than 
60~km~s$^{-1}$. 
} 
\citep{Peimbert78}. 
The Ne, S, and Ar to O ratios do not show any obvious abundance 
anomaly with respect to other PNe \citep{KHM03,HKB04}.  
We emphasize that if the higher electron temperature derived by 
\citet{Eracleous_etal02} were to be used, then the helium abundances 
will increase by 5\%.  
Furthermore, the He$^+$/H$^+$ abundances implied by the line strength 
of He~{\sc i} $\lambda$5876 \AA\ for the bipolar lobes is also high, 
0.20$\pm$0.03.  
We are thus confident on the determination of the helium 
abundances of Kn\,26.

\begin{table}
\caption{Intrinsic Line Intensity Ratios}
\label{tab.flux}
\centering
\begin{tabular}{lrcc}
\hline\hline
\multicolumn{1}{c}{Line ID} & 
\multicolumn{1}{c}{$f(\lambda)$} & 
\multicolumn{1}{c}{Central Ring} & 
\multicolumn{1}{c}{Bipolar Lobes} \\
\hline
$\lambda$3726+3729 [O~{\sc ii}]  &    0.322 &   335$\pm$12  & 495$\pm$40 \\
$\lambda$3869 [Ne~{\sc iii}]   &    0.291 &  49.2$\pm$3.3 &  $\dots$ \\
$\lambda$3889 H8+He~{\sc i}    &    0.286 &  23.0$\pm$1.9 &  $\dots$ \\
$\lambda$3970 [Ne~{\sc iii}]+H$\epsilon$ &    0.203 &  26.63$\pm$1.91 &  $\dots$ \\
$\lambda$4069+76 [S~{\sc ii}]  &    0.238 &  14.8$\pm$2.0 &  $\dots$ \\
$\lambda$4101 H$\delta$        &    0.230 &  24.6$\pm$1.4 &  $\dots$ \\
$\lambda$4340 H$\gamma$        &    0.157 &  49.9$\pm$2.0 &  $\dots$ \\
$\lambda$4471 He~{\sc i}       &    0.115 &   4.5$\pm$0.6 &  $\dots$ \\
$\lambda$4686 He~{\sc ii}      &    0.050 &  26.0$\pm$1.0 &  $\dots$ \\
$\lambda$4861 H$\beta$         &    0.000 & 100.0$\pm$2.2 & 100.0$\pm$3.6 \\
$\lambda$4959 [O~{\sc iii}]    & $-$0.020 & 130.7$\pm$2.6 & 106.7$\pm$3.7 \\
$\lambda$5007 [O~{\sc iii}]    & $-$0.038 &   403$\pm$7   &   319$\pm$9   \\
$\lambda$5198+5200 [N~{\sc i}] & $-$0.104 &   3.9$\pm$0.4 &  $\dots$ \\
$\lambda$5755 [N~{\sc ii}]     & $-$0.131 &   3.2$\pm$0.4 &  $\dots$ \\
$\lambda$5876 He~{\sc i}       & $-$0.203 &  17.8$\pm$0.7 &   28$\pm$4 \\
$\lambda$6300 [O~{\sc i}]      & $-$0.263 &  52.0$\pm$1.9 &   92$\pm$5 \\
$\lambda$6364 [O~{\sc i}]      & $-$0.271 &  16.1$\pm$1.0 &   38$\pm$4 \\
$\lambda$6548 [N~{\sc ii}]     & $-$0.296 &  72.9$\pm$2.2 &    91$\pm$4   \\
$\lambda$6563 H$\alpha$        & $-$0.298 &   285$\pm$8   &   285$\pm$13  \\
$\lambda$6584 [N~{\sc ii}]     & $-$0.300 &   238$\pm$7   &   275$\pm$12  \\
$\lambda$6678 He~{\sc i}       & $-$0.313 &   6.6$\pm$0.4 &  $\dots$ \\
$\lambda$6716 [S~{\sc ii}]     & $-$0.318 &  41.9$\pm$1.5 &  47.4$\pm$2.4 \\
$\lambda$6731 [S~{\sc ii}]     & $-$0.320 &  35.2$\pm$1.3 &  32.8$\pm$1.8 \\
$\lambda$7065 He~{\sc i}       & $-$0.364 &   8.8$\pm$0.5 &  $\dots$ \\
$\lambda$7136 [Ar~{\sc iii}]   & $-$0.374 &  16.0$\pm$0.7 &  $\dots$ \\        
%
%
\hline
\end{tabular}
\end{table}

\begin{table}
\caption{Physical Conditions and Abundances of the Central Ring} 
\label{tab.chem}
\centering
\begin{tabular}{lc}
\hline\hline
\multicolumn{1}{l}{Parameter} & 
\multicolumn{1}{c}{Value} \\
\hline
$T_e$ [N~{\sc ii}] & 9900$\pm$660 K \\
$N_e$ [S~{\sc ii}] & 360$\pm$100 cm$^{-3}$ \\
\hline
$N$(He$^+$)/$N$(H$^+$)    & 0.130$\pm$0.005 \\
$N$(He$^{++}$)/$N$(H$^+$) & 0.030$\pm$0.002 \\
$N$(O$^0$)/$N$(H$^+$)     & (1.0$\pm$0.3)$\times$10$^{-5}$ \\
$N$(O$^+$)/$N$(H$^+$)     & (1.4$\pm$0.5)$\times$10$^{-4}$ \\
$N$(O$^{++}$)/$N$(H$^+$)  & (1.5$\pm$0.4)$\times$10$^{-4}$ \\
$N$(N$^0$)/$N$(H$^+$)     & (5.4$\pm$2.7)$\times$10$^{-6}$ \\
$N$(N$^+$)/$N$(H$^+$)     & (4.7$\pm$1.0)$\times$10$^{-5}$ \\
$N$(S$^+$)/$N$(H$^+$)     & (1.9$\pm$0.4)$\times$10$^{-6}$ \\
$N$(Ar$^{++}$)/$N$(H$^+$) & (1.5$\pm$0.3)$\times$10$^{-6}$ \\
$N$(Ne$^{++}$)/$N$(H$^+$) & (5.2$\pm$1.9)$\times$10$^{-5}$ \\
\hline
He/H  & 0.160$\pm$0.005 \\
O/H   & (3.1$\pm$0.8)$\times$10$^{-4}$ \\
N/H   & (1.1$\pm$0.5)$\times$10$^{-4}$ \\
S/H   & (1.4$\pm$0.5)$\times$10$^{-5}$ \\
Ar/H  & (3.0$\pm$0.9)$\times$10$^{-6}$ \\
Ne/H  & (1.3$\pm$0.8)$\times$10$^{-4}$ \\
\hline
N/O   & 0.34$\pm$0.18 \\
\hline
\end{tabular}
\end{table}

\subsection{The central star of Kn\,26}

\begin{figure}[t]
\begin{center}
\includegraphics[bb=55 215 540 680,width=1.0\linewidth]{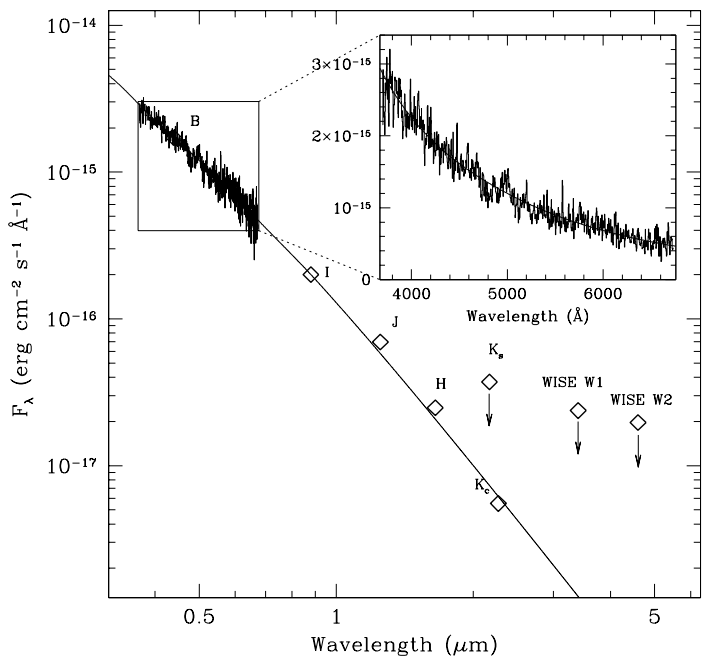}
\caption{
Spectral energy distribution (SED) of the central star of Kn\,26 including 
the optical ALBIREO spectrum (histogram) and broad-band optical, near-IR, 
and \emph{WISE} W1 3.4 $\mu$m and W2 4.6 $\mu$m photometric measurements 
(diamonds).  
The optical ALBIREO spectrum is shown into further detail in 
the inset.  
In both plots, the smooth solid line represents the best fit to the 
optical spectrum by a white dwarf of temperature 70000~K extincted by 
$A_V$=0.65 mag $\equiv$ $c$(H$\beta$)=0.30.  
As discussed in the text, the emission excess in the 2MASS $K_s$ and 
\emph{WISE} W1 and W2 bands with respect to these two fits is due to 
the contribution of nebular emission to these bands, and thus they 
should be regarded as upper limits of the stellar emission.  
}
   \label{fig.sed}  
\end{center}
\end{figure}

The star Lan\,384 is detected in all narrow-band images 
of Kn\,26 inside the elliptical ring-like structure at 
its center (Fig.~\ref{img1}).  
Its optical ALBIREO spectrum and spectral energy distribution (SED) 
additionally including available optical and 2MASS and \emph{WISE} 
IR photometric measurements (Figure~\ref{fig.sed}) show that the 
flux of the star raises bluewards from the near-IR $J$ and $H$ 
bands to the bluest region of the optical spectrum, in agreement with 
\citet{Lanning00} who first recognized Lan\,384 to be a blue star.  
The location of Lan\,384 at the center of Kn\,26 and its 
blue color strongly suggest that it is indeed the central 
star of the PN.  
Paradoxically, the star is not located exactly at the center of 
this ring, as clearly revealed by the insets in the images shown 
in Figure~\ref{img1}.  
We measure a displacement of the central star of Kn\,26 by 
$\approx$0\farcs9 along the direction of the ring's major 
axis at PA $\approx$10\degr.

The 2MASS $K_s$ and \emph{WISE} W1 (3.4 $\mu$m) and W2 (4.6 $\mu$m) 
bands imply an obvious near-IR excess in the SED of Lan\,384.  
An inspection of these images, however, reveals that these 
photometric measurements are contaminated by extended nebular 
emission.  
Using 2MASS $K_s$ photometric measurements of the stars in the field of 
view, we have calibrated our narrow-band $K$ continuum image and derived 
for Lan\,384 a flux density in this band $\approx6$ times lower than that 
implied from the 2MASS $K_s$ magnitude.  
Contrary to the 2MASS $K_s$ photometric measurement, the flux 
density in the $K_c$ filter follows a similar decline to that 
shown by the 2MASS $J$ and $H$ bands.

The available data can be used to estimate the effective 
temperature of Lan\,384.  
In the spectral range covered by the SED in Figure~\ref{fig.sed}, the 
spectrum of the central star of a PN can be well described by a simple 
black-body model.  
Adopting a color excess of $E(B-V)=0.2$ consistent with the optical 
extinction of the nebular spectrum derived in the previous section 
\citep[but also by][]{Eracleous_etal02} and the extinction law of 
\citet{CCM89}, the effective temperature of a black-body that best 
fits the optical spectrum is $\sim$70000~K (solid line in the inset 
of Fig.~\ref{fig.sed}).  
This model also provides a reasonable description of the 
photometric measurements in the $B$, $I$, $J$, $H$, and 
$K_c$ bands, and its temperature is consistent with the 
detection of the nebular He~{\sc ii} $\lambda$4686 \AA\ 
emission line in the nebula implying that about 25\% of 
helium is doubly ionized in the central regions of Kn\,26 
(Tab.~\ref{tab.chem}), which requires effective temperatures 
$\gtrsim$60000~K \citep[e.g.,][]{Pottasch84}.  
We note, however, that this temperature should be regarded as a rough 
estimate because the limited coverage in the blue region of the spectrum 
used to carry out this fit.  
Dedicated UV and high-resolution optical spectrophotometric observations 
of Lan\,384 would be greatly valuable to determine more reliably its 
effective temperature.

\section{Discussion}

The spectroscopic information, excitation, presence of a hot central 
star, and morphology and physical structure clearly confirm the nature 
of PN of the nebula Kn\,26.  
Therefore, we propose its identification as PN\,G084.7--08.0 
following the standard rules of nomenclature for these objects.

The morphological subclass of quadrupolar PNe was introduced by 
\citet{Manchado_etal96} to describe objects which show one single 
equatorial waist and two pair of bipolar lobes with symmetry axes 
oriented along different directions on the plane of the sky.  
Originally this subclass included K\,3-24, M\,1-75, and M\,2-46, and very 
likely M\,3-28 and M\,4-14.  
Since then, the sample of quadrupolar PNe has increased with time 
up to a number of ten 
\citep[][this paper]{Manchado_etal96,GM98,CP00,Mampaso_etal06,Vazquez08,Hsia10},  
%
but they are certainly more because some PNe are prone to be classified 
as quadrupolar \citep[e.g., NGC\,4361 and NGC\,6072,][]{MA01,Kwok_etal10}, 
whereas other morphological subclasses are closely related 
\citep[e.g., the Starfish Nebulae,][]{Sahai00}.  
To date, only one proto-PN, IRAS\,19475+3119, has been reported to have 
a quadrupolar morphology \citep{Sahai07}.

The different orientation of the two pairs of bipolar lobes in 
quadrupolar PNe has been kinematically confirmed to occur also 
along the line of sight for M\,2-46 \citep{Manchado_etal96}, 
NGC\,6881 \citep{GM98}, NGC\,6309 \citep{Vazquez08}, and M\,1-75 
\citep{Santander-Garcia10}.  
The change in the direction of the symmetry axis immediately suggests 
the rotation of the engine collimating the bipolar outflow that shapes 
the bipolar lobes. 
Since this change in direction can be naturally ascribed to the precession 
of a binary system, quadrupolar PNe have been considered archetypes of PNe 
formed after the evolution of the central star in a binary system 
\citep{Manchado_etal96}.

In a review of the properties of a sample of quadrupolar PNe, 
\citet{Mampaso_etal06} concluded that there is little direct 
evidence of detection of binarity among these sources, which 
is otherwise a common problem for the search of binarity among 
PNe \citep{DeMarco09}.  
One possible exception is the central star of IPHAS\,PN-1, whose 
near-IR excess provides tantalizing evidence of a binary system 
\citep{Mampaso_etal06}.  
%
%
%
The 2MASS $K_s$ and \emph{WISE} W1 and W2 photometric measurements of 
Lan\,384 suggest a near-IR excess (Fig.~\ref{fig.sed}), but a careful 
examination of the images led us to conclude that these photometric 
data are contaminated by extended nebular emission.  
A more accurate determination of the star flux using our $K$ 
continuum image confirms that its emission level is consistent 
with the Rayleigh-Jeans tail of a black-body spectral distribution.  
Intriguingly, the central star of Kn\,26 is clearly misplaced 
with respect to the center of the ring-like feature.  
Central stars displaced from the center along the minor axis of the shell are 
observed in many PNe \citep[e.g., MyCn\,18 and Hu\,2-1;][]{Sahai99,Miranda01} 
and can be interpreted as evidence for a binary central star \citep{SRH98}.

\citet{Mampaso_etal06} also investigated the nebular abundances of their 
sample of quadrupolar PNe.  
They concluded that these sources show a great variety of chemical 
abundances that generally do not match the predictions for the 
surface chemical enrichment of single central stars \citep{Marigo_etal03}.  
The nebular chemical abundances of Kn\,26 neither match those 
predictions.  
The abundances of oxygen, nitrogen, neon, and other heavy elements 
are consistent with those of the Sun and the solar neighborhood 
\citep{Asplund_etal09,NP12}, i.e., they do not seem to reflect a 
peculiar chemical enrichment.  
Moreover, the N/O ratio is low, $\approx$0.34, typical of type II PNe.  
In contrast, the helium abundances of Kn\,26 are relatively 
high, with He/H$\approx$0.15, which is more common among type 
I PNe.  
Low values of the N/O ratio and high helium abundances are not 
typically seen in PNe, even in those exhibiting a bipolar 
morphology \citep{Stan_etal06}, but exceptions can be found in 
the literature \citep[e.g., the sample of PNe towards the Galactic 
Bulge described by][]{EC01}.  
Symbiotic stars use to present extremely high helium abundances and low N/O 
ratios \citep[e.g.,][]{LC05} as the increased mass-loss caused by the binary 
companion interactions can curtail the dredge-up of carbon and nitrogen to 
the envelope, affecting the surface chemical enrichment \citep{Lu_etal08}.  
We propose that similar processes may have occurred in Kn\,26, 
resulting in the shortening of the AGB evolution of its 
progenitor and in the rapid stripping of the stellar envelope 
to show helium rich regions. 

The two pairs of bipolar lobes are interwoven in such a way that the small 
bipolar lobes can be described as a protuberance of the surface of the major 
bipolar lobes.  
This significant difference in size between the two pairs of bipolar 
lobes of Kn\,26 seems to imply some time-lap between them, even though 
they have very similar kinematical ages. 
We can envisage the large bipolar lobes forming first, and then, short 
afterwards, a bipolar ejection along a different direction would have 
blown sections of the inner regions of the large bipolar lobes to 
create the minor bipolar lobes.  
The inner lobes would have initially expanded into a medium already 
evacuated by the large bipolar lobes, but then they have interacted 
with the large bipolar lobes walls, resulting in the brightening of 
the emission and distorted velocity field at the tips of the minor 
bipolar lobes.  
At any rate, the time-lap between the ejection of each pair of bipolar 
lobes is presumably much shorter than the kinematical age of the lobes, 
i.e., 1100$\times d$ yr.  
There are other quadrupolar PNe \citep[e.g., M\,1-75,][]{Santander-Garcia10} 
where the two pairs of bipolar lobes formed in a simultaneous ejection.  
This should also be certainly the case for the proto-PN IRAS\,19475+3119, 
as its young age and similar size of the bipolar lobes necessarily imply a 
small time-lap between the two pairs of bipolar lobes \citep{Sahai07}.  
On the other hand, there are quadrupolar PNe 
\citep[e.g., M\,2-46,][]{Manchado_etal96} for 
which the time-lap between ejections can reach 
up to a few thousand years.

\section{Conclusions}

We have used optical and near-IR narrow-band images and optical 
intermediate- and high-dispersion spectroscopic observations to 
investigate the physical structure and chemical abundances of 
Kn\,26.  
The morphological and kinematical information gathered by these observations 
reveal that Kn\,26 is a quadrupolar PN, i.e., it has two pairs of bipolar 
lobes.

The two pairs of bipolar lobes have very similar kinematical ages, 
although the larger size of the major bipolar lobes and the evidence 
of interaction at the tips of the minor bipolar lobes indicate that 
the latter formed during a second bipolar ejection.  
This second ejection was probably close in time to the first one 
that formed the major bipolar lobes, implying a rapid change of 
the referential direction of the collimating mechanism.

The chemical abundances of Kn\,26 are unusual, with a N/O ratio typical 
of type II PNe, but a high helium abundance, typical of type I PNe.  
These chemical abundances cannot be easily reproduced by models 
of single star evolution, but seem to be typical of symbiotic stars.  
We suggest that a companion star could indeed shorten the AGB evolution of 
the progenitor star of Kn\,26 and produce the anomalous chemical abundances 
after going through a common envelope phase, however, no evidence for a 
companion star is provided by the optical and IR SED of the central star.

The comparison of Kn\,26 with other quadrupolar PNe implies a wide 
variety of properties.  
The time-lap between the ejection of the two pairs of bipolar lobes 
may be short, almost coeval, or larger than the dynamical age of the 
bipolar lobes, of a few 1000 yr.  
The chemical abundances are also very different among the members of 
this group, suggesting different progenitors or evolutionary paths.  
These results confirm previous conclusions that the subclass of 
quadrupolar PNe is a rich, not-simple phenomenon.

\begin{acknowledgements}

MAG, LFM and GR-L are partially funded by grant AYA2008-01934 of the Spanish 
Ministerio de Ciencia e Innovaci\'on (MICINN), which includes FEDER funds. 
RV, MAG, and GR-L thank support by grant IN109509 (PAPIIT-DGAPA-UNAM). 
MAG also acknowledges support of the grant AYA 2011-29754-C03-02, and 
LFM acknowledges partial support from grant AYA2011-30228-C03.01 of the
Spanish MINECO, and by grant IN845B-2010/061 of Xunta de Galicia, all of
them partially funded by FEDER funds.
GR-L acknowledges support from CONACyT (grant 177864) and PROMEP (Mexico).  
Finally, we would like to thank the OAN-SPM staff and the CATT for 
time allocation, L.\ Olgu\'\i n for fruitful discussion and assistance 
in the use of the ANNEB package, and an anonymous referee whose 
comments helped us in the analysis and interpretation of the nebular 
spectrum of Kn\,26.  

\end{acknowledgements}

\end{document}